\documentclass[structabstract]{aa}  
\usepackage{graphicx}
\usepackage[usenames,dvipsnames]{color}

\usepackage{txfonts}
\usepackage{longtable}
\usepackage{supertabular}
\usepackage{natbib}
\defcitealias{Fadda+08}{Paper 0}
\begin{document}

\newcommand{\mic}{$\mu$m$\,$}
\newcommand{\mica}{$\mu$m}
\newcommand{\lir}{$\rm{L}_{\rm{IR}} \,$}
\newcommand{\lira}{$\rm{L}_{\rm{IR}}$}
\newcommand{\tsfr}{$\Sigma$(SFR)$\,$}
\newcommand{\tsfra}{$\Sigma$(SFR)}
\newcommand{\tsfrm}{$\Sigma$(SFR)/M$\,$}
\newcommand{\tsfrma}{$\Sigma$(SFR)/M}

\title{The evolution of the star formation activity per halo mass up to redshift $\sim 1.6$ as seen by Herschel
\thanks{Herschel is an ESA space observatory with science instruments provided by 
European-led Principal Investigator consortia and with important 
participation from NASA.}}
\author{P. Popesso\inst{1}, A. Biviano\inst{2}, G. Rodighiero\inst{3}, I. Baronchelli\inst{3}, M. Salvato\inst{1}, A. Saintonge\inst{1}, A. Finoguenov\inst{1}, B. Magnelli\inst{1}, C. Gruppioni\inst{4}, F. Pozzi\inst{5}, D. Lutz\inst{1}, D. Elbaz\inst{6}
\and B. Altieri\inst{7}
\and P. Andreani\inst{8,2}
\and H. Aussel\inst{6}
\and S. Berta\inst{1}
\and P. Capak\inst{9,10}
\and A. Cava\inst{11}
\and A. Cimatti\inst{5}
\and D.~Coia\inst{7}
\and E. Daddi\inst{6}
\and H. Dannerbauer\inst{6}
\and M.~Dickinson\inst{12}
\and K.~Dasyra\inst{6}
\and D. Fadda\inst{13}
\and N. F{\"o}rster Schreiber\inst{1}
\and R. Genzel\inst{1}
\and H.S.~Hwang\inst{6}
\and J.~Kartaltepe\inst{12}
\and O. Ilbert  \inst{14}
\and E. Le Floch\inst{6}
\and R.~Leiton\inst{6}
\and G. Magdis\inst{6}
\and R. Nordon\inst{1}
\and S. Patel \inst{15}
\and A. Poglitsch\inst{1}
\and L. Riguccini\inst{6}
\and M. Sanchez Portal\inst{7} 
\and L. Shao\inst{1}
\and L. Tacconi\inst{1}
\and A. Tomczak\inst{16}
\and K. Tran\inst{16}
\and I. Valtchanov\inst{7}
}
\offprints{Paola Popesso, popesso@mpe.mpg.de}

\institute{Max-Planck-Institut f\"{u}r Extraterrestrische Physik (MPE), Postfach 1312, 85741 Garching, Germany.
\and INAF/Osservatorio Astronomico di Trieste, via G.B. Tiepolo 11, Trieste (Italy) I-34143 
\and Dipartimento di Astronomia, Universit{\`a} di Padova, Vicolo dell'Osservatorio 3, 35122 Padova, Italy.
\and INAF-Osservatorio Astronomico di Bologna, via Ranzani 1, I-40127 Bologna, Italy.
\and Dipartimento di Astronomia, Universit{\`a} di Bologna, Via Ranzani 1, 40127 Bologna, Italy.
\and Laboratoire AIM, CEA/DSM-CNRS-Universit{\'e} Paris Diderot, IRFU/Service d'Astrophysique,  B\^at.709, CEA-Saclay, 91191 Gif-sur-Yvette Cedex, France.
\and Herschel Science Centre, European Space Astronomy Centre, ESA, Villanueva de la Ca\~nada, 28691 Madrid, Spain
\and ESO, Karl-Schwarzschild-Str. 2, D-85748 Garching, Germany.
\and Spitzer Science Center, 314-6 Caltech, 1201 E. California Blvd. Pasadena, 91125
\and Department of Astronomy, 249-17 Caltech, 1201 E. California Blvd. Pasadena, 91125
\and Departamento de Astrof\'{\i}sica, Facultad de CC. Fisicas, Universidad Complutense de Madrid, E28040, Madrid, Spain
\and  National Optical Astronomy Observatory, 950 North Cherry Avenue, Tucson, AZ 85719, USA
\and NASA Herschel Science Center, Caltech 100-22, Pasadena, CA 91125, USA
\and Institute for Astronomy 2680 Woodlawn Drive Honolulu, HI 96822-1897, USA 
\and Leiden Observatory, P.O. Box 9513 NL-2300 RA  Leiden, The Netherlands
\and George P. and Cynthia W. Mitchell Institute for Fundamental Physics and Astronomy, Department of Physics and Astronomy, Texas A\&M University, College Station, TX 77843 
}

\date{Received / Accepted}

\abstract{}{Star formation in massive galaxies is quenched at some
  point during hierarchical mass assembly. To understand where and
  when the quenching processes takes place, we study the evolution of
  the total star formation rate per unit total halo mass (\tsfrma) in
  three different mass scales: low mass halos (field galaxies),
  groups, and clusters, up to a redshift $\rm{z} \approx 1.6$.}{We use
  deep far-infrared PACS data at 100 and 160 \mic to accurately
  estimate the total star formation rate of the Luminous Infrared
  Galaxy population of 9 clusters with mass $\sim 10^{15}$
  $\rm{M}_{\odot}$, and 9 groups/poor clusters with mass $\sim 5
  \times 10^{13}$ $\rm{M}_{\odot}$.  Estimates of the field \tsfrm are
  derived from the literature, by dividing the star formation rate
  density by the mean comoving matter density of the universe.}{The
  field \tsfrm increases with redshift up to $\rm{z} \sim 1$ and it is
  constant thereafter. The evolution of the \tsfrma--z relation in
  galaxy systems is much faster than in the field. Up to redshift
  $\rm{z} \sim 0.2$, the field has a higher \tsfrm than galaxy groups
  and galaxy clusters. At higher redshifts, galaxy groups and the
  field have similar \tsfrma, while massive clusters have
  significantly lower \tsfrm than both groups and the field. There is a hint of a reversal of
the SFR activity vs. environment at $\rm{z} \sim 1.6$, where the
group \tsfrm lies above the field \tsfrm--z relation. We
  discuss possible interpretations of our results in terms of the
  processes of downsizing, and star-formation quenching.}{}


\titlerunning{Evolution of total star formation rate}
\authorrunning{Popesso et al.}

\maketitle

\section{Introduction}
\label{s:intro}
The level of star formation (SF) activity in galaxy systems is known
to be suppressed relative to the field. According to the well known
morphology-density relation \citep{Dressler80} and the star formation
rate (SFR)-density relation \citep{Hashimoto+98,Lewis+02,Gomez+03}, in
the local Universe high density regions, like groups and clusters,
host mostly early type galaxies characterized by a lower level of SF
activity than field (mostly late-type) galaxies.

The environmental dependence of galaxy SFR may change with redshift,
as galaxies in systems undergo significant evolution. In
higher-redshift clusters, the fraction of blue galaxies is higher
\citep[the so-called 'Butcher-Oemler' effect][]{BO78,BO84,Pimbblet03},
and so are the fractions of cluster galaxies with spectra
characterized by young stellar populations
\citep{DG83,Dressler+99,Poggianti+99} and the fraction of galaxies
with late-type morphology
\citep{Dressler+97,Fasano+00,Postman+05,Smith+05}.  Higher-z clusters
are also observed to contain a higher fraction of infrared (IR)
emitting galaxies \citep[the so called ``IR Butcher-Oemler
  effect''][]{STH08,Haines+09b,TDI09}, where most of the IR emission
is powered by SF.

Given this evolution and the morphology- and SFR-density relations
observed locally, it is clear that a quenching of the star formation
activity of galaxies in dense environments is required since $\rm{z}
\sim 1$--2, and that this quenching process must act faster in galaxy
systems than in the field. When and where this quenching process takes
place, is still a matter of debate. SF quenching is generally assumed
to occur within the cluster environment, where processes like ram
pressure \citep{GG72}, cumulative galaxy-galaxy
hydrodynamic/gravitational interactions \citep{Park+09}, strangulation
\citep{LTC80}, and galaxy harassment \citep{Moore+96} are particularly
effective. Indeed, \citet{Tran+09} observed a higher fraction of star
forming galaxies in a supergroups at $z=0.37$ with respect to clusters
at the same redshift. It has however also been claimed that SF
quenching of cluster galaxies occur in low-mass groups or large-scale
filaments prior to cluster assembly \citep[the so-called
  ``pre-processing''][]{ZM98let,
  Kodama+01,Fadda+08,Porter+08,Balogh+11}.  The accretion process of
groups onto clusters can itself lead to star-formation quenching
\citep{Poggianti+04}, perhaps caused by rapid gas consumption due to a
sudden enhancement of the SF activity
\citep{MO03,Coia+05b,Ferrari+05}.  It is also not clear whether
environment plays any role at all in the quenching process. According
to \citet{Peng+10}, mass quenching is the dominant quenching process
for massive galaxies, which generally reside in massive halos like
groups and clusters. Since mass quenching should occur when a galaxy
reaches a limiting mass, more massive galaxies, which reside in more
massive halos, should be quenched earlier than less massive ones,
having reached earlier the limiting mass for quenching. This creates
an environment dependent quenching of galaxy SF, which is driven by
some internal, rather than external, process, such as AGN
feedback. This is supported observationally by the analysis of the
star-formation histories of galaxies in the Virgo cluster region
\citep{Gavazzi+02}, galaxies of higher H-band luminosities being
characterized by shorter timescales of SF. Additional support to this
scenario comes from the analysis of chemical abundances in elliptical
galaxies by \citet{PM04}.

Another way of looking at the evolution of the SF activity in galaxy
systems is to consider a global quantity such as the star formation
rate per unit of halo mass, that is the sum of the SFRs of all the
galaxy in a system, divided by the system total mass, \tsfrma.
According to recent results
\citep{Kodama+04,FZM04,Finn+05,Geach+06,Bai+09,Chung+10,Koyama+10,Hayashi+11}
, the evolution of the cluster \tsfrm is rapid, with a redshift
dependence \mbox{\tsfrma~$\propto (1+\rm{z})^{\alpha}$}, with $\alpha \simeq
5-7$.  The cluster \tsfrm evolves faster up to $z \sim 1$ than the fraction of
IR-emitting galaxies (the IR Butcher-Oemler effect) because
IR-emitting galaxies are not only more numerous in higher-z clusters
but also more IR luminous, given the evolution of the cluster IR
luminosity function \citep{Bai+09}. In addition, there are some
indications that \tsfrm anti-correlates with halo mass in galaxy
systems at similar redshift
\citep{FZM04,Finn+05,Homeier+05,Koyama+10}.

The quantity \tsfrm thus appears to be a powerful tool for analyzing
the rapid evolution of galaxies in systems of different mass.  To
determine \tsfrm one needs to estimate the system dynamical mass
and the SFRs of its member galaxies.  Robust SFR estimates can be
obtained from galaxy IR luminosities, \lira, and these estimates are considered to be
more reliable than the SFR based on optical emission-lines such as
[OII] and H$\alpha$ \citep{Geach+06}.

If one considers only the systems where the total SFR has been
derived from IR observations, the \tsfrma--z relation is currently based
on $\sim 15$ galaxy clusters only \citep{Biviano+11}, all with
masses\footnote{The cluster virial mass, $\rm{M}_{200}$, is the mass
  contained within a sphere of radius $\rm{r}_{200}$, which is the
  radius within which the enclosed average mass density is 200 times
  the critical density.  The two quantities are related by the
  expression $\rm{M}_{200} \equiv 100 \, \rm{H(z)}^2
  \rm{r}_{200}^3/\rm{G}$, where G is the gravitational constant and
  H(z) the Hubble constant at the system redshift.}  $\rm{M}_{200}
\ga 10^{14} \, \rm{M}_{\odot}$ and $\rm{z} < 1$.  Masses for these
systems have been derived in different ways (from X-ray or lensing
data, or from kinematical analyses of the population of cluster
galaxies), and at different limiting radii.

The aim of this paper is to extend the study of the \mbox{\tsfrma--z}
relation to systems of lower masses than clusters, and to higher z,
with a homogeneous estimate of the system masses and total SFRs.
Specifically, we extend the mass range down to the group mass regime,
and to the typical dark matter halo mass of field galaxies
($10^{11}-10^{12} \, \rm{M}_{\odot}$).  We also extend the redshift
range up to $z \sim 1.6$.  We use deep far-infrared PACS data at 100
and 160 \mic to accurately estimate the total SFRs of 18 systems with
masses in the range $2 \times 10^{13} - 3\times \, 10^{15}
\rm{M}_{\odot}$, down to \lira$=10^{11} \rm{L}_{\odot}$, i.e. the
luminosity that characterizes the so-called Luminous IR Galaxies
(LIRGs). This limit is set by the faintest luminosity observable with
our data at z$\sim$1.6.  We use nearly complete galaxy spectroscopic
samples to measure the system masses from their kinematics.  To extend
the \tsfrma-z relation to the field regime, i.e.  to dark matter halo
masses typical of field galaxies, we use published global SFR
densities at different redshifts
\citep{Madau+98,Gruppioni+11,Magnelli+11}.

In Sect.~\ref{s:sample} and Sect.~\ref{s:specphot} we describe our
sample and our spectroscopic and photometric data-set, in
Sect.~\ref{s:members} we describe how we define system membership of
the galaxies with available redshifts, and derive the system dynamical
masses, in Sect.~\ref{s:sfr} we determine the total SFRs of the
systems.  The resulting \tsfrma-z relations are presented in
Sect.~\ref{s:res} for clusters, groups (Sect.~\ref{s:cls}), and the
field (Sect.~\ref{s:field}). We summarize and discuss our results in
Sect.~\ref{s:conc}.

We adopt H$_0=70$ km~s$^{-1}$~Mpc$^{-1}$, $\Omega_m=0.3$,
$\Omega_{\Lambda}=0.7$ throughout this paper. We adopt a Salpeter IMF
when calculating stellar masses and SFRs.

\section{The data-set}
\label{s:data}
\subsection{The sample of galaxy systems}
\label{s:sample}

Our sample comprises 9 X-ray detected systems at $0.15 < z < 0.85$,
all observed in the PACS Evolutionary Probe PACS GT Program
\citep[PEP,][]{Lutz+11}.  The systems are listed in Table~\ref{t1}. In
addition, we use the X-ray detected group and cluster catalog of the
COSMOS field (Finoguenov et al., in prep.) to identify 27 systems
at $z < 0.8$ with at least 10 spectroscopic members each, the minimum
required for an acceptable mass estimate from the system kinematics
\citep{Girardi+93}. The COSMOS field is observed as part of the PEP
program.  We further include in our sample 4 systems in the GOODS
North and South fields at $0.7 < z < 1.6$ observed as part of the PACS
GOODS-Herschel Program (Elbaz et al. 2011). Of these systems, two are
X-ray detected, one at $z=0.73$ in \mbox{GOODS-S} \citep{Gilli+03,LeFevre+04,Popesso+09}, another
one at $z=1.02$ in \mbox{GOODS-N} \citep{Elbaz+07}, one system lies close to
the Chandra CCD chip gap and it is not X-ray detected \citep[\mbox{GOODS-N}
  system at $z=0.85$][]{Bauer+02}, and the highest redshift system (at
$z=1.612$) is only marginally X-ray detected in the Chandra GOODS-S
map \citep{Kurk+08}.  Finally, we include in our sample the Bullet
cluster \citep{Barrena+02,Markevitch+04}, a well-known massive cluster
undergoing a violent collision with an infalling group, which we use
as a test-case for the dependence of the \tsfrm on the cluster
dynamical state. PACS data for the Bullet cluster are not public
yet. For this cluster we use the \lir based on Spitzer MIPS 24 \mic
data published by \citet{Chung+10}.  The considered LIRGs members lie
within 1.7 Mpc from the cluster center.  This is close to the
$\rm{r}_{200}$ radius, $\sim 2$~Mpc, estimated from the kinematics of
cluster galaxies \citep{Barrena+02}.

We split our sample in groups and clusters depending on their mass
estimates (see next section), groups/poor clusters with $10^{13}
\, \rm{M}_{\odot} < \rm{M}_{200} < 3 \times 10^{14} \, \rm{M}_{\odot}$ and
clusters with $\rm{M}_{200} \geq 3 \times 10^{14} \, M_{\odot}$. The mass
value separating groups/poor clusters from clusters approximately
corresponds to the richness class 0 in the catalog of \citet{ACO89}
\citep[using the relations between mass and velocity dispersion and
  between velocity dispersion and richness class for galaxy
  clusters; see][]{Girardi+93,Biviano+06}.

\subsection{Spectral and photometric data}
\label{s:specphot}
We take the data for the COSMOS field from the multi-wavelength and
IRAC-selected catalog of \citet{Ilbert+10} complemented with public
Spitzer 24 \mic \citep{LeFloch+09,Sanders+07} and PACS 100 and 160
\mic data \citep{Lutz+11}.  The association between 24 \mic and PACS
sources with their optical counterparts is done via a maximum likelihood
method \citep[see][for details]{Lutz+11,Berta+10}. The photometric
sources were cross-matched in coordinates with the sources for which a
high-confidence spectroscopic redshift is available. For this purpose
we use the public catalogs of spectroscopic redshifts complemented
with other unpublished data. This catalog includes redshifts from
either SDSS or the public zCOSMOS-bright data acquired using VLT/VIMOS
\citep{Lilly+07,Lilly+09} complemented with Keck/DEIMOS (PIs:
Scoville, Capak, Salvato, Sanders, Kartaltepe), Magellan/IMACS
\citep{Trump+07}, and MMT \citep{Prescott+06} spectroscopic redshifts.

Similar multi-wavelength (from UV to 24 \mica) data for the GOODS-S
field is provided by the MUSIC catalog \citep{Grazian+06,Santini+09}
and complemented with PACS data at 70, 100 and 160 \mic and
spectroscopic redshifts from the PEP catalog (see Lutz et al. 2011 for
more details) and from the GMASS survey \citep{Cimatti+08}.

We take the data for the GOODS-N field from the multi-wavelength
catalog prepared by the PEP Team \citep{Berta+10} including UV to
MIPS 24 data, PACS 100 and 160 \mic data, complemented with the
spectroscopic redshift catalog of \citet{BCW08}.

We collect publicly available spectroscopic data for all clusters from
the NED. For CL 0024$+$17 and MS 0451$-$03 we also collect
multi-wavelength data from \citet{Moran+05,Moran+07}. Multi-wavelength
photometric data are not available for the other clusters.  The
references for each cluster spectroscopic sample are listed in Table~\ref{t1}.
Each cluster sample is complemented with MIPS 24 \mic data, when
available, and PACS 100 and 160 \mic data
\citep{Lutz+11}.

The data reduction, source extraction, and a detailed explanation of
the depths of the PACS maps used in this paper can be found in
\citet{Lutz+11} and \citet{Elbaz+11}, for the PEP and GOODS-Herschel
surveys, respectively.

\begin{table*}
\centering
\begin{tabular}{ l  l  l  c   r  r  r  r}
\hline
\hline
name                            & z~~      & $\rm{r}_{200}$~ & $\rm{M}_{200}$~~~               & $N_m$~ & $N_{LIRG}$ & \%~$z_{spec}$~ & $\rm{ref}$ \\
                                     &        &  (Mpc)     & ($10^{14} M_{\odot}$)&           &                          &                      &   \\
\hline
 (1)                                  & (2)         &  (3)                     & (4)                                          &    (5)         &   (6)            &    (7)                  & (8)  \\
\hline
\hline
\multicolumn{8}{c}{Cluster sample} \\
\hline
A2218                          & 0.175 &  2.73     & 27.6 & 107 & 1 & 75\% & 1\\
A2219                          & 0.226 &  2.375   & 19    &  126 & 2 & 68\% & 2\\
A2390                          & 0.228 &  2.533   & 23     &  259 & 3 & 69\% & 3\\
MS 1358$+$62            &  0.329 & 1.711   & 8.2    & 286 & 1 & 73\% & 4\\
$\rm{COSMOS\_CL1}^a$  & 0.351  & 1.2    & 4.8     & 36 & 6 & 48\% & 5 \\
MS 0451$-$03            & 0.539  & 1.966   & 15.5 & 342 & 8 & 80\% & 6\\
$\rm{COSMOS\_CL2}^a$  & 0.78   & 1.18   & 3.88  & 36 & 6 & 48\% & 5 \\
MS 1054$-$03             & 0.830 & 1.521  & 10.1 &  134 & 5 & 100\% & 7\\
RX J0152$-$13            &  0.838 & 1.657 &13.3 & 228 & 11 & 96\% & 8\\
\hline
\hline
\multicolumn{8}{c}{Group/poor cluster sample} \\
\hline
$\rm{COSMOS\_GR1}^b$ & 0.1     & 0.61   & 0.26  &   47  & 1 & 56\% & 5\\
$\rm{COSMOS\_GR2}^b$ & 0.22   & 0.62  & 0.31   &  77 & 1 & 66\% & 5\\
$\rm{COSMOS\_GR3}^b$ & 0.35   & 0.68   & 0.31  &   81 & 4 & 48\% & 5\\
CL 0024$+$17             & 0.395 & 0.96   & 1.5     & 396 & 10 & 92\% & 6,9\\
$\rm{COSMOS\_GR4}^b$  & 0.70  & 0.69   & 0.31  &   34    & 10 & 47\% & 5 \\ 
$\rm{GOODS-S\_01}$  & 0.735 & 0.51   & 0.85   & 188 & 3 & 88\% & 10 \\
$\rm{GOODS-N\_01}$ & 0.85   & 0.51   & 0.9     & 112 & 3 & 90\% & 11\\
$\rm{GOODS-N\_02}$ & 1.02   & 0.73   & 1.03  &  130 & 4 & 79\% & 11\\
$\rm{GOODS-S\_02}$  & 1.61   & 0.52   & 0.9    & 76 & 6 & 68\% & 10\\
\hline
\hline
\multicolumn{8}{c}{Merging system} \\
\hline
Bullet cluster                & 0.297 & 1.7       & 9.5     & 43    & 5 & --     & 12 \\
\hline
\hline
\end{tabular}
\caption{The table lists the following main properties of the cluster
  and group samples: column (1): name; column (2): redshift; column (3):
  radius $\rm{r}_{200}$ in Mpc; column (4): mass $\rm{M}_{200}$ in $10^{14} \, \rm{M}_{\odot}$;
  column (5): number of members used in the estimate of $\rm{M}_{200}$;
  column (6): number of LIRG members within $\rm{r}_{200}$; column (7):
  spectroscopic completeness of the LIRG sample within $\rm{r}_{200}$;
  column (8): references for spectroscopic data: 1 \citet{LeBorgne+92}; 
2 SDSS DR7 \citep{Abazajian+09}; 3 \citet{Yee+96}; 
4 \citet{Fisher+98}; 5 COSMOS
  spectroscopic sample, see text; 6 \citet{Moran+05}; 7 \citet{Tran+07};
  8 \citet{Patel+09}; 9 \citet{Moran+07}; 10 PEP-GOODS-S
  spectrophotometric catalog \citet{Popesso+11}; 11 PEP-GOODS-N
  spectrophotometric catalog \citet{Berta+10}; 12 \citet{Chung+10}.}
\label{t1}
\begin{minipage}{1.0\hsize}
\tiny
a Composite of COSMOS clusters. The quantities reported for this system are the mean of the contributing systems.\\
b Composite of COSMOS groups. The quantities reported for this system are the mean of the contributing systems. \\
\end{minipage}
\end{table*}

\subsection{Membership and mass estimates of galaxy systems}
\label{s:members}

To define the galaxies that are members of the systems described in
Sect.~\ref{s:sample}, we adopt the algorithm of \citet{Mamon+10},
which is based on the modeling of the mass and anisotropy profiles of
cluster-sized halos extracted from a cosmological numerical
simulation. This algorithm is more effective than traditional
approaches \citep[e.g.][]{YV77} in rejecting interlopers, while still
preserving cluster members.  The system membership selection depends
on the location of galaxies in the system-centric distance --
rest-frame velocity\footnote{The galaxy rest-frame velocities with
  respect to the system mean velocity are obtained by the usual
  relation \mbox{$\rm{v} = c (\rm{z}-\overline{\rm{z}}) / (1+
    \overline{\rm{z}})$} \citep{HN79}, where $\overline{\rm{z}}$ is
  the system mean redshift, determined with the biweight estimator
  \citep{BFG90}.} diagram. The peaks of the X-ray surface brightness
are adopted as centers of the X-ray detected systems. The galaxy
number density maxima (estimated using an adaptive kernel technique)
are adopted as centers for the other systems.  The interloper
rejection procedure is iterated until convergence.

\begin{figure}
\begin{center}
\begin{minipage}{0.5\textwidth}
\resizebox{\hsize}{!}{\includegraphics{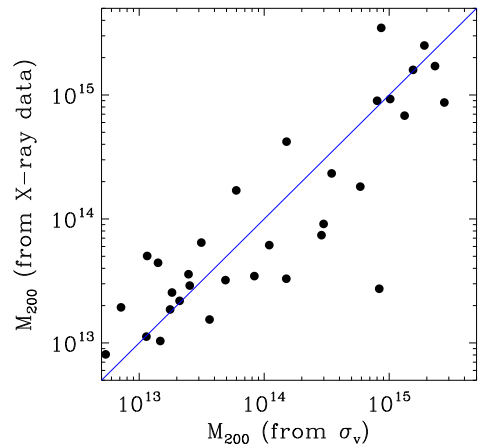}}
\end{minipage}
\end{center}
\caption{Comparison of system mass estimates obtained from X-ray data with those obtained from system velocity dispersions.}
\label{fig1}
\end{figure}

As an outcome of the procedure of \citet{Mamon+10}, the system
dynamical masses are obtained from biweight estimates \citep{BFG90}
of the system velocity dispersions
along the line-of-sight, $\sigma_{\rm{v}}$, as in \citet[][see
  Appendix A]{MM07}. We use the $\sigma_{\rm{v}}$-based mass
estimates, rather than the more traditional virial mass estimates
because the $\sigma_{\rm{v}}$-based mass estimates are robust against
problems of spatial incompleteness, unlike the more traditional virial
theorem estimator \citep[see][]{Biviano+06}, so we do not have to worry
about the possibility that in some systems of our sample the spectroscopic
completeness is different in different regions.

The uncertainties in the system mass estimates are estimated from the
uncertainties in the biweight estimate of $\sigma_{\rm{v}}$
\citep{BFG90} via the propagation of error analysis.

As an alternative mass estimate we consider that obtained by assuming
hydrostatic equilibrium of the X-ray-emitting intra-cluster or
intra-group gas. The two mass-estimates are compared in
Fig.~\ref{fig1} for those systems which have both estimates
available. The scatter is large, but there is no significant bias, on
average, between the two mass estimates. There is only one system
    for which the two mass estimates are in disagreement by more than
    $3\sigma$. The discrepancy is probably due the low number of
    spectroscopic members (9) available for this system. We exclude it
    from our sample for further analyses.

\subsection{Star formation rates}
\label{s:sfr}
 
We fit the spectral energy distributions (SEDs) of all system members
in our sample which have multiwavelength data available from the UV to
the PACS wavelengths, using a set of empirical templates of local
objects \citep{Polletta+07,Gruppioni+10} that reproduce
most of the observed galaxy SEDs from the optical to the FIR
wavelengths \citep[see][]{Rodighiero+10}.  We then determine the
galaxy \lir by integrating the best-fit SED models from 8
to 1000 \mica. 

We estimate the \lir of
those system members for which multiwavelength data are not available,
by fitting their PACS data with SED
models from \citet{Polletta+07}, to extract the ${\nu} \rm{L}_{\nu}$ at 100
\mic rest frame, and by using the \lira~$ - {\nu} \rm{L}_{\nu}$ relation, as
explained in Popesso et al. (in prep.).

Some of our member galaxies do not have PACS data either due to the
incompleteness of the PACS photometric catalog between the 3 and
5$\sigma$ levels, or because the PACS catalog are not deep enough to
reach the required $10^{11} \rm{L}_{\odot}$ limits at the redshift of the
system.  We determine the \lir of these galaxies from their MIPS 24
\mic flux densities by using the scaling relation of \citet{CE01},
corrected as explained in Popesso et al. (in prep.)  to avoid
overestimations \citep{Elbaz+11,Nordon+11}.  We also re-determine in
the same way the \lir of the member galaxies of the Bullet cluster
from the 24 \mic data of \citet{Chung+10}. In total, only 8 galaxies
(9\% of our member LIRGs) do not have their \lir estimated from PACS
data: 2 in MS 0451$-$03, 4 in RX J0152$-$13 and 2 in the $z\sim 1.6$
GOODS-S group.

The main advantage of using PACS data in the determination of the \lir
is to remove possible contamination by the emission of active galactic
nuclei (AGN), since most of the rest-frame far-IR emission comes from
the host galaxy \citep{Netzer+07,Lutz+10}. Thus, we do not need to
exclude AGN hosts from our sample as was instead done in previous
similar analysis \citep[e.g.][]{Geach+06,Bai+09,Biviano+11}. AGN
contamination is a concern, however, when estimating the SFR from 24
$\mu$m data. We do exclude the AGN hosts from the evaluation of the
total SFR of the Bullet cluster, using the AGN classification of
\citet{Chung+10}, since PACS data are not yet publicly available for
this cluster. We also cross-correlated the system members, whose \lir
is based on 24 \mic data only, with publicly available AGN catalogs
\citep{Johnson+03,Demarco+05,Brusa+09}. None of those members turn
out to be an AGN.

Finally, we estimate the galaxy SFRs from their \lir via the law of
Kennicutt (1999). The LIRG \lira-limit corresponds to a limiting SFR of
17 M$_{\odot}$~yr$^{-1}$.

We estimate \tsfra, the total SFR of each system, by summing up the
SFRs of the LIRGs that are members of the system.  We then correct
\tsfr for spectroscopic incompletetness in the following way. For all
systems where the PACS data reach the $10^{11} \rm{L}_{\odot}$ limit at the system
redshift, we take as a reference the PACS 100 \mic catalog. We plot
the 100 \mic flux density vs. \lir of system member galaxies in narrow
redshift bins to estimate the 100 \mic flux density that corresponds
to the LIRG \lir at each redshift, f$_{\rm{LIRG}}(\rm{z})$.  The
spectroscopic incompleteness is defined as the fraction of PACS
sources with z, among all those with flux $>\rm{f}_{\rm{LIRG}}(\rm{z})$
within $\rm{r}_{200}$.  The inverse of this fraction is the correction
factor that we apply to the \tsfr estimates to correct for
incompleteness.  For the systems which contain member galaxies for
which the \lira-estimates are based on MIPS 24 \mic flux densities, we
use the same approach by using the 24 \mic catalog as reference.

To test the reliability of this method, we consider a system with
100\% spectroscopic completeness down to the LIR \lir level,
MS~1054$-$03 \citep{Bai+09}. We perform 500000 random extractions from
the luminosity function of MS~1054$-$03 of a number of LIRGs
corresponding to a given spectroscopic completeness.  We repeat this
test for all the spectroscopic coverage values listed in Table 1. We
then estimate the uncertainty due to the incompleteness correction
from the dispersion of the difference between the recovered \tsfr and
the real value. The uncertainty due to this correction is $\simeq
20$\% in the range of spectroscopic completeness of our sample.

The total uncertainty of the \tsfr estimates is determined from the
propagation of error analysis, by considering a 10\% uncertainty in
the \lir estimates \citep[see][for further details]{Lutz+11}, and the
20\% uncertainty due to the completeness correction.

The spectroscopic coverage of the COSMOS field is much lower than
those of the clusters and GOODS fields and spatially not
uniform. Thus, to reduce the error bars of the \tsfra , we combine the
COSMOS groups and, separately, the COSMOS clusters in four, and,
respectively, two redshift bins, see Table~\ref{t1}.  In each z-bin we then
define the mean \tsfra, $\rm{M}_{200}$, and spectroscopic
incompleteness, separately for the groups and the clusters
contributing to the sample in that bin.

Counting the composite COSMOS groups and clusters as individual
systems, our final sample comprises 9 massive clusters, 9 groups/poor
clusters, and the Bullet cluster (see Table~\ref{t1}).

\begin{figure}
\begin{center}
\begin{minipage}{0.5\textwidth}
\resizebox{\hsize}{!}{\includegraphics{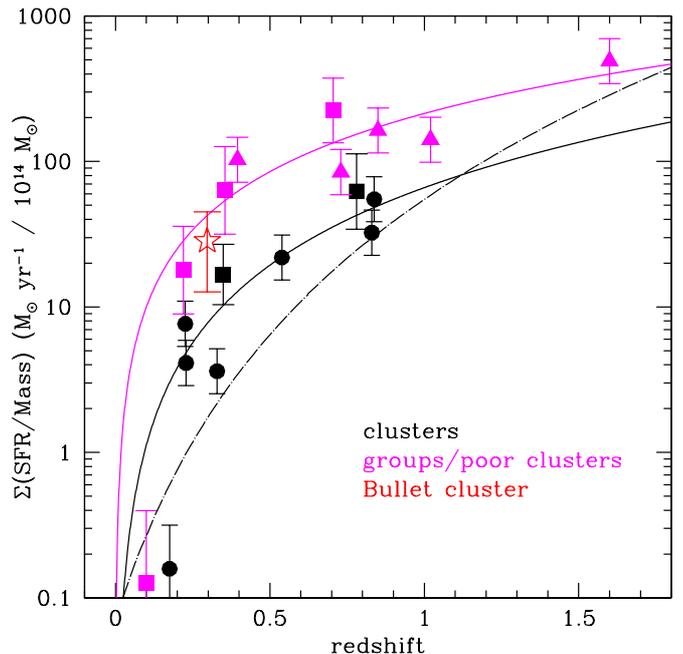}}
\end{minipage}
\end{center}
\caption{\tsfrma- redshift relation for clusters (black symbols) and
  groups (magenta symbols). Square symbols, triangles, and dots
  identify, respectively, the COSMOS composite systems, the GOODS
  systems, and the remaining systems. The red star identifies the
  Bullet cluster.  The black solid line shows the best fit \tsfrma--z
  relation for the cluster sample, excluding the Bullet cluster. The relation fitted to the data is of the type \tsfrma~$\propto z^{\alpha}$. The
  dashed line shows the relation, \tsfrma~$\propto (1+z)^{\alpha}$,  of \citet{Bai+09} rescaled to the
  LIRG \lira-regime and to the $\rm{r}_{200}$ region.  The magenta
  solid lines shows the best fit for the sample of groups and poor
  clusters. }
\label{fig2}
\end{figure}

\section{Results: the \tsfrma--z relation}
\label{s:res}
\subsection{Clusters versus groups}
\label{s:cls}

The \tsfrma--redshift relation is shown in Fig. \ref{fig2}.  Black,
and magenta symbols show the \tsfrma - redshift relation for the
clusters and, respectively, the groups. For both samples there is
evidence for a significant \tsfrm vs. z correlation, 99\% significant
according to a Spearman rank correlation test.  On the other hand, we
do not find evidence for significant \mbox{\tsfrma-$\rm{M}_{200}$} or
$\rm{M}_{200}$-z correlations within the group and cluster
samples separately. Hence the observed \tsfrma--z relation within each
sample must be interpreted as a genuine redshift evolution of the
\tsfrm of galaxy systems.

Groups appear to be characterized by higher \tsfrm values than
clusters, at all redshifts, i.e. they show a higher SF
activity than massive clusters. In this sense, when the sample of
groups and clusters are considered together, \tsfrm does anticorrelate
with the system $\rm{M}_{200}$, as already found in previous
studies \citep{FZM04,Finn+05,Homeier+05,Koyama+10}.

The solid curves in Fig.~\ref{fig2} represent best-fit models to the
observed \tsfrma--z relations, \tsfrm$=(66\pm 23) \times
\rm{z}^{1.77\pm0.36}$ for the cluster sample and \tsfrm$=(213\pm 44)
\times \rm{z}^{1.33\pm0.34}$ for the group sample.  In Fig.~\ref{fig2}
we also show as a dashed line the \tsfrm$\propto (1+\rm{z})^{5.3}$
model proposed by \citet{Bai+09}.  We rescale their relation, which is
built for \tsfr computed down to a SFR limit of 2
M$_{\odot}$~yr$^{-1}$, to our adopted LIRG limit, by using the IR
luminosity function of \citet{Bai+09}. Moreover, the \tsfr in Bai et
al.'s relation is evaluated within a radius of $\rm{r}_{500}$, the
radius corresponding to an overdensity equal to 500 times the critical
density. To convert to our adopted radius $\rm{r}_{200}$, we adopt a
\citet{NFW97} model, wich concentration $\mbox{c=5}$ to scale for the number
of galaxies inside the two radii and derive the
$\rm{r}_{500}/\rm{r}_{200}$ ratio, and then use Fig.~7 in
\citet{Bai+09} to account for the different fraction of IR-emitting
galaxies at $\rm{r}_{500}$ and $\rm{r}_{200}$.  The modified relation
of \citet{Bai+09} appears to fail to predict the rapid evolution of
cluster \tsfrma, as already suggested by \citet{Biviano+11}.

\subsection{Structures versus field}
\label{s:field}

\begin{figure}
\begin{center}
\begin{minipage}{0.5\textwidth}
\resizebox{\hsize}{!}{\includegraphics{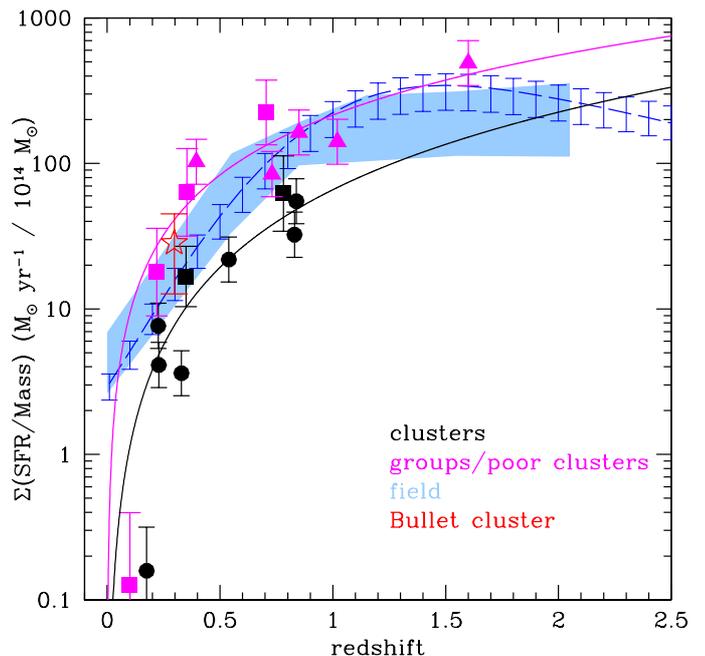}}
\end{minipage}
\end{center}
\caption{The field \tsfrma-redshift relation from \citet{Magnelli+11}
  (light blue shaded region) and \citet{Gruppioni+11} (dashed blue
  line). The shading and error bars represent 1$\sigma$
    confidence levels  Other symbols/lines have the same meaning of
  the symbols as in Fig. \ref{fig2}}
\label{fig3}
\end{figure}

In Fig. \ref{fig3} we compare the \tsfrma--z relation of galaxy systems
with the corresponding relation for field galaxies (light blue shaded region and
blue dashed line). The \tsfrma--z relation of field galaxies is obtained by
dividing the observed Star Formation Rate Density (SFRD) of
\citet[][triangles]{Magnelli+11} and the modeled SFRD of
\citet[][dashed line]{Gruppioni+11}, by the mean comoving density of
the universe  ($\Omega_m \times
\rho_c $ where $\Omega_m=0.3$ and $\rho_c$ is the
critical density of the Universe). Both SFRD have been
evaluated only down to the SFR corresponding to the LIRG \lir,
via the Kennicutt relation.

The field SFRD has been estimated in large comoving volumes that
include galaxy systems, voids, and isolated galaxies, and is thus
representative of the general field galaxy population.  According to
the dark halo mass function of \citet{Jenkins+01}, halos of
$10^{11}-10^{12}$ $\rm{M}_{\odot}$ give the main contribution to the
dark matter budget at all redshifts. Thus, the \tsfrm of
Fig.~\ref{fig3} can be considered as an effective estimate of the
\tsfrm of galaxy-sized dark matter halos.

Observed and modeled \tsfrm are in very good agreement within the
error bars.  They increase from $\rm{z}=0$ to $\rm{z} \sim 1$ where
they reach a plateau.

The field \tsfrma--z relation lies above both the group and the
cluster relations at $\rm{z} \la 0.2$. The field \tsfrm is more than
an order of magnitude higher than those of Abell~2218 at $z=0.175$ and
of the composite COSMOS group at $z=0.1$.  At higher redshifts, group
and field galaxy halos show comparable \tsfrma. Possibly, the group
\tsfrm lie above the field relation at $\rm{z} \sim 1.6$, consistently
with a reversal of the SFR- density relation observed by \citet{Elbaz+07},
\citet{Cooper+08}, and \citet{Popesso+11}. However,
this conclusion is based on a single group \tsfrm determination.
Clusters have lower \tsfrm than field galaxy halos at all redshifts up
to the last measured point at $\rm{z}=0.85$. A blind extrapolation of
the best-fit cluster \tsfrma--z relation would suggest that clusters
should display a higher SFR per unit mass than field galaxies at at $z
\ga 2$. However, such an extrapolation is extremely uncertain with the
present data and massive systems as the ones considered here should
not even exist at such redshift according to $\Lambda$CDM hierarchical
model predictions.

\section{Discussion and conclusions}
\label{s:conc}

We find significant evidence for an evolution of the \tsfrm with z in
clusters, groups, and the field. Both the cluster and the group
\tsfrm increase monotonically with z and  the field \tsfrm reaches a
maximum at $\rm{z} \ga 1$.

The cluster \tsfrm evolution we find, appears to be faster, at least
in the range $0 < \rm{z} < 1$, compared to the evolutionary relation
suggested by \citet{Geach+06,Bai+07,Bai+09,Koyama+10}. This might be
due to our higher-\lir cut, if LIRGs evolve faster than less luminous
IR-emitting galaxies. A faster evolution has already been suggested by
\citet{Biviano+11}, but only in the $\rm{z} \la 0.5$ range, while in
our data there is no sign of a flattening of the \tsfrma--z relation
for clusters above that redshift.

We confirm earlier indications \citep{Finn+05,Bai+07,Koyama+10} that
\tsfrm is lower for systems of higher mass, and we show for the first
time that this is true at all redshifts from $\rm{z} \sim 0$ to $\rm{z}
\sim 0.9$. This is true when we compare clusters to groups and the
field.  On the other hand, the group \tsfrm lies below the field
relation only at $\rm{z} \la 0.2$. At higher redshifts, the group and
the field \tsfrm become comparable. There is a hint of a reversal of
the SFR activity vs. environment at $\rm{z} \sim 1.6$, where the
group \tsfrm lies above the field \tsfrm--z relation.

Our results appear to support a scenario in which the quenching of SF
occurs earlier in galaxies embedded in more massive halos, i.e. first
in clusters, then in groups, and finally in the field. This would be
consistent with a ``halo downsizing'' effect, whereby massive halos
evolve more rapidly than low mass halos \citep{NvdBD06}. Although in a
narrow range of halo masses, the ``halo downsizing'' effect has
already been observed in the stellar-to-total mass ratio vs. redshift
relation by \citet{Foucaud+10}.



What is causing the difference between the cluster and the group/field
\tsfrm--z relation?  The similarity of the field and the groups
\tsfrm--z relations suggests that a SF quenching effect is taking
place mostly after galaxies enter the cluster environment, and not in
groups before they merge into more massive structures, as would be
predicted by the pre-processing scenario
\citep{ZM98let,Kodama+01,Balogh+11}. We can therefore consider the
field and group \tsfrm--z relation as indicating the normal galaxy
evolution, and the cluster \tsfrm--z relation as the accelerated
evolution experienced by galaxies following their accretion to more
massive structures, as individuals or in groups. However, we point out
that other differences may intervene between the star forming
population of the groups and the field. Indeed, in agrement with our
results, \citet{Tran+09} find consistent levels of star formation in
the supergroup at $z=0.37$ and the field population, but they also
reveal a much higher fraction of early type galaxies among the SF
galaxies of the group environment with respect to the field. Similar
results are obtained by \citet{HP09}. This could indicate a
morphological transformation happening prior to the SF quenching
effect.

Part of the observed evolution may come from internal, rather than
external processes. In fact, most of the LIRGs in our sample are
likely to be rather massive galaxies \citep[as inferred from the
  almost linear, z-dependent, relation between galaxy SFRs and stellar
  masses of][]{Elbaz+07}, and according to \citet{Peng+10} massive
galaxies evolve mostly because of an internally driven process, called
'mass quenching', caused perhaps by feedback from active galactic
nuclei. But since this process is unlikely to be more efficient in
quenching SF of massive galaxies in clusters than in other
environments, we need to consider an additional quenching process that
would operate only in the cluster environment.

Two obvious candidates for producing this additional quenching, are
ram-pressure stripping \citep{GG72} and starvation \citep{LTC80}.
Ram-pressure stripping is a fast process \citep{AMB99} and is
effective only in galaxy systems where the gas density is high,
i.e. in clusters.  However, ram-pressure stripping seems ruled out by
the presence of LIRGs contributed by the infalling group (the
``bullet'') in the Bullet cluster. The group LIRG contribution is
probably the reason of the position of the Bullet cluster in the
\tsfrma--z plot, intermediate between that of clusters and groups
\citep{Chung+10}.  If rapid ram-pressure stripping was a major star
formation quenching process, the group LIRGs could have had their star
formation quenched already, given that the cluster-group collision
happened $\sim 250$ Myr ago \citep{SF07}.

Starvation, caused by the removal of the hot gas halo reservoirs of
galaxies, is a more likely candidate given that we see an
accelerated evolution (relative to the field) both in clusters and in
groups \citep{KM08,Bekki09}. The removal of galaxy hot gas reservoirs
inducing starvation can be caused by tidal galaxy-galaxy encounters or
by the interaction with the intra-cluster/intra-group medium.
Starvation should proceed more effectively in higher (galaxy or gas)
density regions, hence it should quench SF earlier in cluster than 
in group galaxies, as we observe. 

In order to better constrain these evolutionary scenarios it is
important to extend the analysis of the \tsfrm--z relation to even
higher redshifts, where we might expect to see a reversal of the
environment dependence as we approach the epoch when galaxies in dense
environments underwent their first major episodes of SF. In the near
future we plan to use new Herschel PACS observations of 6 poor and
rich clusters at $1.4 < \rm{z} < 1.9$ to explore this issue.

\begin{acknowledgements}
PACS has been developed by a consortium of institutes led by MPE 
(Germany) and including UVIE (Austria); KUL, CSL, IMEC (Belgium); CEA, 
OAMP (France); MPIA (Germany); IFSI, OAP/AOT, OAA/CAISMI, LENS, SISSA 
(Italy); IAC (Spain). This development has been supported by the funding 
agencies BMVIT (Austria), ESA-PRODEX (Belgium), CEA/CNES (France),
DLR (Germany), ASI (Italy), and CICYT/MCYT (Spain).

We gratefully acknowledge the contributions of the entire COSMOS
collaboration consisting of more than 100 scientists. More information
about the COSMOS survey is available at
http://www.astro.caltech.edu/$\sim$cosmos.

This research has made use of NASA's Astrophysics Data System, of NED,
which is operated by JPL/Caltech, under contract with NASA, and of
SDSS, which has been funded by the Sloan Foundation, NSF, the US
Department of Energy, NASA, the Japanese Monbukagakusho, the Max
Planck Society, and the Higher Education Funding Council of England.
The SDSS is managed by the participating institutions
(www.sdss.org/collaboration/credits.html).

\end{acknowledgements}

\bibliography{master}

\begin{thebibliography}{107}
\expandafter\ifx\csname natexlab\endcsname\relax\def\natexlab#1{#1}\fi

\bibitem[{{Abadi} {et~al.}(1999){Abadi}, {Moore}, \& {Bower}}]{AMB99}
{Abadi}, M.~G., {Moore}, B., \& {Bower}, R.~G. 1999, \mnras, 308, 947

\bibitem[{{Abazajian} {et~al.}(2009){Abazajian}, {Adelman-McCarthy},
  {Ag{\"u}eros}, {Allam}, {Allende Prieto}, {An}, {Anderson}, {Anderson},
  {Annis}, {Bahcall}, \& et~al.}]{Abazajian+09}
{Abazajian}, K.~N., {Adelman-McCarthy}, J.~K., {Ag{\"u}eros}, M.~A., {et~al.}
  2009, \apjs, 182, 543

\bibitem[{{Abell} {et~al.}(1989){Abell}, {Corwin}, \& {Olowin}}]{ACO89}
{Abell}, G.~O., {Corwin}, Jr., H.~G., \& {Olowin}, R.~P. 1989, \apjs, 70, 1

\bibitem[{{Bai} {et~al.}(2007){Bai}, {Marcillac}, {Rieke}, {Rieke}, {Tran},
  {Hinz}, {Rudnick}, {Kelly}, \& {Blaylock}}]{Bai+07}
{Bai}, L., {Marcillac}, D., {Rieke}, G.~H., {et~al.} 2007, \apj, 664, 181

\bibitem[{{Bai} {et~al.}(2009){Bai}, {Rieke}, {Rieke}, {Christlein}, \&
  {Zabludoff}}]{Bai+09}
{Bai}, L., {Rieke}, G.~H., {Rieke}, M.~J., {Christlein}, D., \& {Zabludoff},
  A.~I. 2009, \apj, 693, 1840

\bibitem[{{Balogh} {et~al.}(2011){Balogh}, {McGee}, {Wilman}, {Finoguenov},
  {Parker}, {Connelly}, {Mulchaey}, {Bower}, {Tanaka}, \&
  {Giodini}}]{Balogh+11}
{Balogh}, M.~L., {McGee}, S.~L., {Wilman}, D.~J., {et~al.} 2011, \mnras, 412,
  2303

\bibitem[{{Barger} {et~al.}(2008){Barger}, {Cowie}, \& {Wang}}]{BCW08}
{Barger}, A.~J., {Cowie}, L.~L., \& {Wang}, W.-H. 2008, \apj, 689, 687

\bibitem[{{Barrena} {et~al.}(2002){Barrena}, {Biviano}, {Ramella}, {Falco}, \&
  {Seitz}}]{Barrena+02}
{Barrena}, R., {Biviano}, A., {Ramella}, M., {Falco}, E.~E., \& {Seitz}, S.
  2002, \aap, 386, 816

\bibitem[{{Bauer} {et~al.}(2002){Bauer}, {Alexander}, {Brandt},
  {Hornschemeier}, {Miyaji}, {Garmire}, {Schneider}, {Bautz}, {Chartas},
  {Griffiths}, \& {Sargent}}]{Bauer+02}
{Bauer}, F.~E., {Alexander}, D.~M., {Brandt}, W.~N., {et~al.} 2002, \aj, 123,
  1163

\bibitem[{{Beers} {et~al.}(1990){Beers}, {Flynn}, \& {Gebhardt}}]{BFG90}
{Beers}, T.~C., {Flynn}, K., \& {Gebhardt}, K. 1990, \aj, 100, 32

\bibitem[{{Bekki}(2009)}]{Bekki09}
{Bekki}, K. 2009, \mnras, 399, 2221

\bibitem[{{Berta} {et~al.}(2010){Berta}, {Magnelli}, {Lutz}, {Altieri},
  {Aussel}, {Andreani}, {Bauer}, {Bongiovanni}, {Cava}, {Cepa}, {Cimatti},
  {Daddi}, {Dominguez}, {Elbaz}, {Feuchtgruber}, {F{\"o}rster Schreiber},
  {Genzel}, {Gruppioni}, {Katterloher}, {Magdis}, {Maiolino}, {Nordon},
  {P{\'e}rez Garc{\'{\i}}a}, {Poglitsch}, {Popesso}, {Pozzi}, {Riguccini},
  {Rodighiero}, {Saintonge}, {Santini}, {Sanchez-Portal}, {Shao}, {Sturm},
  {Tacconi}, {Valtchanov}, {Wetzstein}, \& {Wieprecht}}]{Berta+10}
{Berta}, S., {Magnelli}, B., {Lutz}, D., {et~al.} 2010, \aap, 518, L30

\bibitem[{{Biviano} {et~al.}(2011){Biviano}, {Fadda}, {Durret}, {Edwards}, \&
  {Marleau}}]{Biviano+11}
{Biviano}, A., {Fadda}, D., {Durret}, F., {Edwards}, L.~O.~V., \& {Marleau}, F.
  2011, \aap, 532, A77

\bibitem[{{Biviano} {et~al.}(2006){Biviano}, {Murante}, {Borgani}, {Diaferio},
  {Dolag}, \& {Girardi}}]{Biviano+06}
{Biviano}, A., {Murante}, G., {Borgani}, S., {et~al.} 2006, \aap, 456, 23

\bibitem[{{Brusa} {et~al.}(2009){Brusa}, {Fiore}, {Santini}, {Grazian},
  {Comastri}, {Zamorani}, {Hasinger}, {Merloni}, {Civano}, {Fontana}, \&
  {Mainieri}}]{Brusa+09}
{Brusa}, M., {Fiore}, F., {Santini}, P., {et~al.} 2009, \aap, 507, 1277

\bibitem[{{Butcher} \& {Oemler}(1978)}]{BO78}
{Butcher}, H. \& {Oemler}, Jr., A. 1978, \apj, 219, 18

\bibitem[{{Butcher} \& {Oemler}(1984)}]{BO84}
{Butcher}, H. \& {Oemler}, Jr., A. 1984, \apj, 285, 426

\bibitem[{{Chary} \& {Elbaz}(2001)}]{CE01}
{Chary}, R. \& {Elbaz}, D. 2001, \apj, 556, 562

\bibitem[{{Chung} {et~al.}(2010){Chung}, {Gonzalez}, {Clowe}, {Markevitch}, \&
  {Zaritsky}}]{Chung+10}
{Chung}, S.~M., {Gonzalez}, A.~H., {Clowe}, D., {Markevitch}, M., \&
  {Zaritsky}, D. 2010, \apj, 725, 1536

\bibitem[{{Cimatti} {et~al.}(2008){Cimatti}, {Robberto}, {Baugh}, {Beckwith},
  {Content}, {Daddi}, {De Lucia}, {Garilli}, {Guzzo}, {Kauffmann}, {Lehnert},
  {Maccagni}, {Mart{\'{\i}}nez-Sansigre}, {Pasian}, {Reid}, {Rosati},
  {Salvaterra}, {Stiavelli}, {Wang}, {Osorio}, {Balcells}, {Bersanelli},
  {Bertoldi}, {Blaizot}, {Bottini}, {Bower}, {Bulgarelli}, {Burgasser},
  {Burigana}, {Butler}, {Casertano}, {Ciardi}, {Cirasuolo}, {Clampin}, {Cole},
  {Comastri}, {Cristiani}, {Cuby}, {Cuttaia}, {de Rosa}, {Sanchez}, {di Capua},
  {Dunlop}, {Fan}, {Ferrara}, {Finelli}, {Franceschini}, {Franx}, {Franzetti},
  {Frenk}, {Gardner}, {Gianotti}, {Grange}, {Gruppioni}, {Gruppuso}, {Hammer},
  {Hillenbrand}, {Jacobsen}, {Jarvis}, {Kennicutt}, {Kimble}, {Kriek}, {Kurk},
  {Kneib}, {Le Fevre}, {Macchetto}, {MacKenty}, {Madau}, {Magliocchetti},
  {Maino}, {Mandolesi}, {Masetti}, {McLure}, {Mennella}, {Meyer}, {Mignoli},
  {Mobasher}, {Molinari}, {Morgante}, {Morris}, {Nicastro}, {Oliva},
  {Padovani}, {Palazzi}, {Paresce}, {Garrido}, {Pian}, {Popa}, {Postman},
  {Pozzetti}, {Rayner}, {Rebolo}, {Renzini}, {R{\"o}ttgering}, {Schinnerer},
  {Scodeggio}, {Saisse}, {Shanks}, {Shapley}, {Sharples}, {Shea}, {Silk},
  {Smail}, {Span{\'o}}, {Steinacker}, {Stringhetti}, {Szalay}, {Tresse},
  {Trifoglio}, {Urry}, {Valenziano}, {Villa}, {Perez}, {Walter}, {Ward},
  {White}, {White}, {Wright}, {Wyse}, {Zamorani}, {Zacchei}, {Zeilinger}, \&
  {Zerbi}}]{Cimatti+08}
{Cimatti}, A., {Robberto}, M., {Baugh}, C., {et~al.} 2008, Experimental
  Astronomy, 37

\bibitem[{{Coia} {et~al.}(2005){Coia}, {McBreen}, {Metcalfe}, {Biviano},
  {Altieri}, {Ott}, {Fort}, {Kneib}, {Mellier}, {Miville-Desch{\^e}nes},
  {O'Halloran}, \& {Sanchez-Fernandez}}]{Coia+05b}
{Coia}, D., {McBreen}, B., {Metcalfe}, L., {et~al.} 2005, \aap, 431, 433

\bibitem[{{Cooper} {et~al.}(2008){Cooper}, {Newman}, {Weiner}, {Yan},
  {Willmer}, {Bundy}, {Coil}, {Conselice}, {Davis}, {Faber}, {Gerke},
  {Guhathakurta}, {Koo}, \& {Noeske}}]{Cooper+08}
{Cooper}, M.~C., {Newman}, J.~A., {Weiner}, B.~J., {et~al.} 2008, \mnras, 383,
  1058

\bibitem[{{Demarco} {et~al.}(2005){Demarco}, {Rosati}, {Lidman}, {Homeier},
  {Scannapieco}, {Ben{\'{\i}}tez}, {Mainieri}, {Nonino}, {Girardi}, {Stanford},
  {Tozzi}, {Borgani}, {Silk}, {Squires}, \& {Broadhurst}}]{Demarco+05}
{Demarco}, R., {Rosati}, P., {Lidman}, C., {et~al.} 2005, \aap, 432, 381

\bibitem[{{Dressler}(1980)}]{Dressler80}
{Dressler}, A. 1980, \apj, 236, 351

\bibitem[{{Dressler} \& {Gunn}(1983)}]{DG83}
{Dressler}, A. \& {Gunn}, J.~E. 1983, \apj, 270, 7

\bibitem[{{Dressler} {et~al.}(1997){Dressler}, {Oemler}, {Couch}, {Smail},
  {Ellis}, {Barger}, {Butcher}, {Poggianti}, \& {Sharples}}]{Dressler+97}
{Dressler}, A., {Oemler}, A.~J., {Couch}, W.~J., {et~al.} 1997, \apj, 490, 577

\bibitem[{{Dressler} {et~al.}(1999){Dressler}, {Smail}, {Poggianti}, {Butcher},
  {Couch}, {Ellis}, \& {Oemler}}]{Dressler+99}
{Dressler}, A., {Smail}, I., {Poggianti}, B.~M., {et~al.} 1999, \apjs, 122, 51

\bibitem[{{Elbaz} {et~al.}(2007){Elbaz}, {Daddi}, {Le Borgne}, {Dickinson},
  {Alexander}, {Chary}, {Starck}, {Brandt}, {Kitzbichler}, {MacDonald},
  {Nonino}, {Popesso}, {Stern}, \& {Vanzella}}]{Elbaz+07}
{Elbaz}, D., {Daddi}, E., {Le Borgne}, D., {et~al.} 2007, \aap, 468, 33

\bibitem[{{Elbaz} {et~al.}(2011){Elbaz}, {Dickinson}, {Hwang}, {Diaz-Santos},
  {Magdis}, {Magnelli}, {Le Borgne}, {Galliano}, {Pannella}, {Chanial},
  {Armus}, {Charmandaris}, {Daddi}, {Aussel}, {Popesso}, {Kartaltepe},
  {Altieri}, {Valtchanov}, {Coia}, {Dannerbauer}, {Dasyra}, {Leiton},
  {Mazzarella}, {Buat}, {Burgarella}, {Chary}, {Gilli}, {Ivison}, {Juneau},
  {LeFloc'h}, {Lutz}, {Morrison}, {Mullaney}, {Murphy}, {Pope}, {Scott},
  {Alexander}, {Brodwin}, {Calzetti}, {Cesarsky}, {Charlot}, {Dole},
  {Eisenhardt}, {Ferguson}, {Foerster-Schreiber}, {Frayer}, {Giavalisco},
  {Huynh}, {Koekemoer}, {Papovich}, {Reddy}, {Surace}, {Teplitz}, {Yun}, \&
  {Wilson}}]{Elbaz+11}
{Elbaz}, D., {Dickinson}, M., {Hwang}, H.~S., {et~al.} 2011, 1105.2537

\bibitem[{{Fadda} {et~al.}(2008){Fadda}, {Biviano}, {Marleau},
  {Storrie-Lombardi}, \& {Durret}}]{Fadda+08}
{Fadda}, D., {Biviano}, A., {Marleau}, F.~R., {Storrie-Lombardi}, L.~J., \&
  {Durret}, F. 2008, \apjl, 672, L9

\bibitem[{{Fasano} {et~al.}(2000){Fasano}, {Poggianti}, {Couch}, {Bettoni},
  {Kj{\ae}rgaard}, \& {Moles}}]{Fasano+00}
{Fasano}, G., {Poggianti}, B.~M., {Couch}, W.~J., {et~al.} 2000, \apj, 542, 673

\bibitem[{{Ferrari} {et~al.}(2005){Ferrari}, {Benoist}, {Maurogordato},
  {Cappi}, \& {Slezak}}]{Ferrari+05}
{Ferrari}, C., {Benoist}, C., {Maurogordato}, S., {Cappi}, A., \& {Slezak}, E.
  2005, \aap, 430, 19

\bibitem[{{Finn} {et~al.}(2004){Finn}, {Zaritsky}, \& {McCarthy}}]{FZM04}
{Finn}, R.~A., {Zaritsky}, D., \& {McCarthy}, Jr., D.~W. 2004, \apj, 604, 141

\bibitem[{{Finn} {et~al.}(2005){Finn}, {Zaritsky}, {McCarthy}, {Poggianti},
  {Rudnick}, {Halliday}, {Milvang-Jensen}, {Pell{\'o}}, \& {Simard}}]{Finn+05}
{Finn}, R.~A., {Zaritsky}, D., {McCarthy}, Jr., D.~W., {et~al.} 2005, \apj,
  630, 206

\bibitem[{{Fisher} {et~al.}(1998){Fisher}, {Fabricant}, {Franx}, \& {van
  Dokkum}}]{Fisher+98}
{Fisher}, D., {Fabricant}, D., {Franx}, M., \& {van Dokkum}, P. 1998, \apj,
  498, 195

\bibitem[{{Foucaud} {et~al.}(2010){Foucaud}, {Conselice}, {Hartley}, {Lane},
  {Bamford}, {Almaini}, \& {Bundy}}]{Foucaud+10}
{Foucaud}, S., {Conselice}, C.~J., {Hartley}, W.~G., {et~al.} 2010, \mnras,
  406, 147

\bibitem[{{Gavazzi} {et~al.}(2002){Gavazzi}, {Boselli}, {Pedotti}, {Gallazzi},
  \& {Carrasco}}]{Gavazzi+02}
{Gavazzi}, G., {Boselli}, A., {Pedotti}, P., {Gallazzi}, A., \& {Carrasco}, L.
  2002, \aap, 396, 449

\bibitem[{{Geach} {et~al.}(2006){Geach}, {Smail}, {Ellis}, {Moran}, {Smith},
  {Treu}, {Kneib}, {Edge}, \& {Kodama}}]{Geach+06}
{Geach}, J.~E., {Smail}, I., {Ellis}, R.~S., {et~al.} 2006, \apj, 649, 661

\bibitem[{{Gilli} {et~al.}(2003){Gilli}, {Cimatti}, {Daddi}, {Hasinger},
  {Rosati}, {Szokoly}, {Tozzi}, {Bergeron}, {Borgani}, {Giacconi}, {Kewley},
  {Mainieri}, {Mignoli}, {Nonino}, {Norman}, {Wang}, {Zamorani}, {Zheng}, \&
  {Zirm}}]{Gilli+03}
{Gilli}, R., {Cimatti}, A., {Daddi}, E., {et~al.} 2003, \apj, 592, 721

\bibitem[{{Girardi} {et~al.}(1993){Girardi}, {Biviano}, {Giuricin},
  {Mardirossian}, \& {Mezzetti}}]{Girardi+93}
{Girardi}, M., {Biviano}, A., {Giuricin}, G., {Mardirossian}, F., \&
  {Mezzetti}, M. 1993, \apj, 404, 38

\bibitem[{{G{\'o}mez} {et~al.}(2003){G{\'o}mez}, {Nichol}, {Miller}, {Balogh},
  {Goto}, {Zabludoff}, {Romer}, {Bernardi}, {Sheth}, {Hopkins}, {Castander},
  {Connolly}, {Schneider}, {Brinkmann}, {Lamb}, {SubbaRao}, \&
  {York}}]{Gomez+03}
{G{\'o}mez}, P.~L., {Nichol}, R.~C., {Miller}, C.~J., {et~al.} 2003, \apj, 584,
  210

\bibitem[{{Grazian} {et~al.}(2006){Grazian}, {Fontana}, {de Santis}, {Nonino},
  {Salimbeni}, {Giallongo}, {Cristiani}, {Gallozzi}, \&
  {Vanzella}}]{Grazian+06}
{Grazian}, A., {Fontana}, A., {de Santis}, C., {et~al.} 2006, \aap, 449, 951

\bibitem[{{Gruppioni} {et~al.}(2010){Gruppioni}, {Pozzi}, {Andreani},
  {Rodighiero}, {Cimatti}, {Altieri}, {Aussel}, {Berta}, {Bongiovanni},
  {Brisbin}, {Cava}, {Cepa}, {Daddi}, {Dominguez-Sanchez}, {Elbaz},
  {F{\"o}rster Schreiber}, {Genzel}, {Le Floc'h}, {Lutz}, {Magdis},
  {Magliocchetti}, {Magnelli}, {Maiolino}, {Nordon}, {Per{\'e}z-Garc{\'{\i}}a},
  {Poglitsch}, {Popesso}, {Riguccini}, {Saintonge}, {Sanchez-Portal},
  {Santini}, {Shao}, {Sturm}, {Tacconi}, \& {Valtchanov}}]{Gruppioni+10}
{Gruppioni}, C., {Pozzi}, F., {Andreani}, P., {et~al.} 2010, \aap, 518, L27

\bibitem[{{Gruppioni} {et~al.}(2011){Gruppioni}, {Pozzi}, {Zamorani}, \&
  {Vignali}}]{Gruppioni+11}
{Gruppioni}, C., {Pozzi}, F., {Zamorani}, G., \& {Vignali}, C. 2011, 1105.1955

\bibitem[{{Gunn} \& {Gott}(1972)}]{GG72}
{Gunn}, J.~E. \& {Gott}, J.~R. 1972, \apj, 176, 1

\bibitem[{{Haines} {et~al.}(2009){Haines}, {Smith}, {Egami}, {Ellis}, {Moran},
  {Sanderson}, {Merluzzi}, {Busarello}, \& {Smith}}]{Haines+09b}
{Haines}, C.~P., {Smith}, G.~P., {Egami}, E., {et~al.} 2009, \apj, 704, 126

\bibitem[{{Harrison} \& {Noonan}(1979)}]{HN79}
{Harrison}, E.~R. \& {Noonan}, T.~W. 1979, \apj, 232, 18

\bibitem[{{Hashimoto} {et~al.}(1998){Hashimoto}, {Oemler}, {Lin}, \&
  {Tucker}}]{Hashimoto+98}
{Hashimoto}, Y., {Oemler}, A.~J., {Lin}, H., \& {Tucker}, D.~L. 1998, \apj,
  499, 589

\bibitem[{{Hayashi} {et~al.}(2011){Hayashi}, {Kodama}, {Koyama}, {Tadaki}, \&
  {Tanaka}}]{Hayashi+11}
{Hayashi}, M., {Kodama}, T., {Koyama}, Y., {Tadaki}, K.-I., \& {Tanaka}, I.
  2011, \mnras, 415, 2670

\bibitem[{{Homeier} {et~al.}(2005){Homeier}, {Demarco}, {Rosati}, {Postman},
  {Blakeslee}, {Bouwens}, {Bradley}, {Ford}, {Goto}, {Gronwall}, {Holden},
  {Jee}, {Martel}, {Mei}, {Menanteau}, {Zirm}, {Clampin}, {Hartig},
  {Illingworth}, {Ardila}, {Bartko}, {Ben{\'{\i}}tez}, {Broadhurst}, {Brown},
  {Burrows}, {Cheng}, {Cross}, {Feldman}, {Franx}, {Golimowski}, {Infante},
  {Kimble}, {Krist}, {Lesser}, {Meurer}, {Miley}, {Motta}, {Sirianni},
  {Sparks}, {Tran}, {Tsvetanov}, {White}, \& {Zheng}}]{Homeier+05}
{Homeier}, N.~L., {Demarco}, R., {Rosati}, P., {et~al.} 2005, \apj, 621, 651

\bibitem[{{Hwang} \& {Park}(2009)}]{HP09}
{Hwang}, H.~S. \& {Park}, C. 2009, \apj, 700, 791

\bibitem[{{Ilbert} {et~al.}(2010){Ilbert}, {Salvato}, {Le Floc'h}, {Aussel},
  {Capak}, {McCracken}, {Mobasher}, {Kartaltepe}, {Scoville}, {Sanders},
  {Arnouts}, {Bundy}, {Cassata}, {Kneib}, {Koekemoer}, {Le F{\`e}vre}, {Lilly},
  {Surace}, {Taniguchi}, {Tasca}, {Thompson}, {Tresse}, {Zamojski}, {Zamorani},
  \& {Zucca}}]{Ilbert+10}
{Ilbert}, O., {Salvato}, M., {Le Floc'h}, E., {et~al.} 2010, \apj, 709, 644

\bibitem[{{Jenkins} {et~al.}(2001){Jenkins}, {Frenk}, {White}, {Colberg},
  {Cole}, {Evrard}, {Couchman}, \& {Yoshida}}]{Jenkins+01}
{Jenkins}, A., {Frenk}, C.~S., {White}, S.~D.~M., {et~al.} 2001, \mnras, 321,
  372

\bibitem[{{Johnson} {et~al.}(2003){Johnson}, {Best}, \& {Almaini}}]{Johnson+03}
{Johnson}, O., {Best}, P.~N., \& {Almaini}, O. 2003, \mnras, 343, 924

\bibitem[{{Kawata} \& {Mulchaey}(2008)}]{KM08}
{Kawata}, D. \& {Mulchaey}, J.~S. 2008, \apjl, 672, L103

\bibitem[{{Kodama} {et~al.}(2004){Kodama}, {Balogh}, {Smail}, {Bower}, \&
  {Nakata}}]{Kodama+04}
{Kodama}, T., {Balogh}, M.~L., {Smail}, I., {Bower}, R.~G., \& {Nakata}, F.
  2004, \mnras, 354, 1103

\bibitem[{{Kodama} {et~al.}(2001){Kodama}, {Smail}, {Nakata}, {Okamura}, \&
  {Bower}}]{Kodama+01}
{Kodama}, T., {Smail}, I., {Nakata}, F., {Okamura}, S., \& {Bower}, R.~G. 2001,
  \apjl, 562, L9

\bibitem[{{Koyama} {et~al.}(2010){Koyama}, {Kodama}, {Shimasaku}, {Hayashi},
  {Okamura}, {Tanaka}, \& {Tokoku}}]{Koyama+10}
{Koyama}, Y., {Kodama}, T., {Shimasaku}, K., {et~al.} 2010, \mnras, 403, 1611

\bibitem[{{Kurk} {et~al.}(2008){Kurk}, {Cimatti}, {Zamorani}, {Halliday},
  {Mignoli}, {Pozzetti}, {Daddi}, {Rosati}, {Dickinson}, {Bolzonella},
  {Cassata}, {Renzini}, {Franceschini}, {Rodighiero}, \& {Berta}}]{Kurk+08}
{Kurk}, J., {Cimatti}, A., {Zamorani}, G., {et~al.} 2008, in Astronomical
  Society of the Pacific Conference Series, Vol. 399, Panoramic Views of Galaxy
  Formation and Evolution, ed. {T.~Kodama, T.~Yamada, \& K.~Aoki}, 332

\bibitem[{{Larson} {et~al.}(1980){Larson}, {Tinsley}, \& {Caldwell}}]{LTC80}
{Larson}, R.~B., {Tinsley}, B.~M., \& {Caldwell}, C.~N. 1980, \apj, 237, 692

\bibitem[{{Le Borgne} {et~al.}(1992){Le Borgne}, {Pello}, \&
  {Sanahuja}}]{LeBorgne+92}
{Le Borgne}, J.~F., {Pello}, R., \& {Sanahuja}, B. 1992, \aaps, 95, 87

\bibitem[{{Le F{\`e}vre} {et~al.}(2004){Le F{\`e}vre}, {Vettolani}, {Paltani},
  {Tresse}, {Zamorani}, {Le Brun}, {Moreau}, {Bottini}, {Maccagni}, {Picat},
  {Scaramella}, {Scodeggio}, {Zanichelli}, {Adami}, {Arnouts}, {Bardelli},
  {Bolzonella}, {Cappi}, {Charlot}, {Contini}, {Foucaud}, {Franzetti},
  {Garilli}, {Gavignaud}, {Guzzo}, {Ilbert}, {Iovino}, {McCracken}, {Mancini},
  {Marano}, {Marinoni}, {Mathez}, {Mazure}, {Meneux}, {Merighi}, {Pell{\`o}},
  {Pollo}, {Pozzetti}, {Radovich}, {Zucca}, {Arnaboldi}, {Bondi}, {Bongiorno},
  {Busarello}, {Ciliegi}, {Gregorini}, {Mellier}, {Merluzzi}, {Ripepi}, \&
  {Rizzo}}]{LeFevre+04}
{Le F{\`e}vre}, O., {Vettolani}, G., {Paltani}, S., {et~al.} 2004, \aap, 428,
  1043

\bibitem[{{Le Floc'h} {et~al.}(2009){Le Floc'h}, {Aussel}, {Ilbert},
  {Riguccini}, {Frayer}, {Salvato}, {Arnouts}, {Surace}, {Feruglio},
  {Rodighiero}, {Capak}, {Kartaltepe}, {Heinis}, {Sheth}, {Yan}, {McCracken},
  {Thompson}, {Sanders}, {Scoville}, \& {Koekemoer}}]{LeFloch+09}
{Le Floc'h}, E., {Aussel}, H., {Ilbert}, O., {et~al.} 2009, \apj, 703, 222

\bibitem[{{Lewis} {et~al.}(2002){Lewis}, {Balogh}, {De Propris}, {Couch},
  {Bower}, {Offer}, {Bland-Hawthorn}, {Baldry}, {Baugh}, {Bridges}, {Cannon},
  {Cole}, {Colless}, {Collins}, {Cross}, {Dalton}, {Driver}, {Efstathiou},
  {Ellis}, {Frenk}, {Glazebrook}, {Hawkins}, {Jackson}, {Lahav}, {Lumsden},
  {Maddox}, {Madgwick}, {Norberg}, {Peacock}, {Percival}, {Peterson},
  {Sutherland}, \& {Taylor}}]{Lewis+02}
{Lewis}, I., {Balogh}, M., {De Propris}, R., {et~al.} 2002, \mnras, 334, 673

\bibitem[{{Lilly} {et~al.}(2009){Lilly}, {Le Brun}, {Maier}, {Mainieri},
  {Mignoli}, {Scodeggio}, {Zamorani}, {Carollo}, {Contini}, {Kneib}, {Le
  F{\`e}vre}, {Renzini}, {Bardelli}, {Bolzonella}, {Bongiorno}, {Caputi},
  {Coppa}, {Cucciati}, {de la Torre}, {de Ravel}, {Franzetti}, {Garilli},
  {Iovino}, {Kampczyk}, {Kovac}, {Knobel}, {Lamareille}, {Le Borgne}, {Pello},
  {Peng}, {P{\'e}rez-Montero}, {Ricciardelli}, {Silverman}, {Tanaka}, {Tasca},
  {Tresse}, {Vergani}, {Zucca}, {Ilbert}, {Salvato}, {Oesch}, {Abbas},
  {Bottini}, {Capak}, {Cappi}, {Cassata}, {Cimatti}, {Elvis}, {Fumana},
  {Guzzo}, {Hasinger}, {Koekemoer}, {Leauthaud}, {Maccagni}, {Marinoni},
  {McCracken}, {Memeo}, {Meneux}, {Porciani}, {Pozzetti}, {Sanders},
  {Scaramella}, {Scarlata}, {Scoville}, {Shopbell}, \& {Taniguchi}}]{Lilly+09}
{Lilly}, S.~J., {Le Brun}, V., {Maier}, C., {et~al.} 2009, \apjs, 184, 218

\bibitem[{{Lilly} {et~al.}(2007){Lilly}, {Le F{\`e}vre}, {Renzini}, {Zamorani},
  {Scodeggio}, {Contini}, {Carollo}, {Hasinger}, {Kneib}, {Iovino}, {Le Brun},
  {Maier}, {Mainieri}, {Mignoli}, {Silverman}, {Tasca}, {Bolzonella},
  {Bongiorno}, {Bottini}, {Capak}, {Caputi}, {Cimatti}, {Cucciati}, {Daddi},
  {Feldmann}, {Franzetti}, {Garilli}, {Guzzo}, {Ilbert}, {Kampczyk}, {Kovac},
  {Lamareille}, {Leauthaud}, {Borgne}, {McCracken}, {Marinoni}, {Pello},
  {Ricciardelli}, {Scarlata}, {Vergani}, {Sanders}, {Schinnerer}, {Scoville},
  {Taniguchi}, {Arnouts}, {Aussel}, {Bardelli}, {Brusa}, {Cappi}, {Ciliegi},
  {Finoguenov}, {Foucaud}, {Franceschini}, {Halliday}, {Impey}, {Knobel},
  {Koekemoer}, {Kurk}, {Maccagni}, {Maddox}, {Marano}, {Marconi}, {Meneux},
  {Mobasher}, {Moreau}, {Peacock}, {Porciani}, {Pozzetti}, {Scaramella},
  {Schiminovich}, {Shopbell}, {Smail}, {Thompson}, {Tresse}, {Vettolani},
  {Zanichelli}, \& {Zucca}}]{Lilly+07}
{Lilly}, S.~J., {Le F{\`e}vre}, O., {Renzini}, A., {et~al.} 2007, \apjs, 172,
  70

\bibitem[{{Lutz} {et~al.}(2010){Lutz}, {Mainieri}, {Rafferty}, {Shao},
  {Hasinger}, {Wei{\ss}}, {Walter}, {Smail}, {Alexander}, {Brandt}, {Chapman},
  {Coppin}, {F{\"o}rster Schreiber}, {Gawiser}, {Genzel}, {Greve}, {Ivison},
  {Koekemoer}, {Kurczynski}, {Menten}, {Nordon}, {Popesso}, {Schinnerer},
  {Silverman}, {Wardlow}, \& {Xue}}]{Lutz+10}
{Lutz}, D., {Mainieri}, V., {Rafferty}, D., {et~al.} 2010, \apj, 712, 1287

\bibitem[{{Lutz} {et~al.}(2011){Lutz}, {Poglitsch}, {Altieri}, {Andreani},
  {Aussel}, {Berta}, {Bongiovanni}, {Brisbin}, {Cava}, {Cepa}, {Cimatti},
  {Daddi}, {Dominguez-Sanchez}, {Elbaz}, {Forster Schreiber}, {Genzel},
  {Grazian}, {Gruppioni}, {Harwit}, {Le Floc'h}, {Magdis}, {Magnelli},
  {Maiolino}, {Nordon}, {Perez Garcia}, {Popesso}, {Pozzi}, {Riguccini},
  {Rodighiero}, {Saintonge}, {Sanchez Portal}, {Santini}, {Shao}, {Sturm},
  {Tacconi}, {Valtchanov}, {Wetzstein}, \& {Wieprecht}}]{Lutz+11}
{Lutz}, D., {Poglitsch}, A., {Altieri}, B., {et~al.} 2011, 1106.3285

\bibitem[{{Madau} {et~al.}(1998){Madau}, {Pozzetti}, \& {Dickinson}}]{Madau+98}
{Madau}, P., {Pozzetti}, L., \& {Dickinson}, M. 1998, \apj, 498, 106

\bibitem[{{Magnelli} {et~al.}(2011){Magnelli}, {Elbaz}, {Chary}, {Dickinson},
  {Le Borgne}, {Frayer}, \& {Willmer}}]{Magnelli+11}
{Magnelli}, B., {Elbaz}, D., {Chary}, R.~R., {et~al.} 2011, \aap, 528, A35

\bibitem[{{Mamon} {et~al.}(2010){Mamon}, {Biviano}, \& {Murante}}]{Mamon+10}
{Mamon}, G.~A., {Biviano}, A., \& {Murante}, G. 2010, \aap, 520, A30

\bibitem[{{Markevitch} {et~al.}(2004){Markevitch}, {Gonzalez}, {Clowe},
  {Vikhlinin}, {Forman}, {Jones}, {Murray}, \& {Tucker}}]{Markevitch+04}
{Markevitch}, M., {Gonzalez}, A.~H., {Clowe}, D., {et~al.} 2004, \apj, 606, 819

\bibitem[{{Mauduit} \& {Mamon}(2007)}]{MM07}
{Mauduit}, J.-C. \& {Mamon}, G.~A. 2007, \aap, 475, 169

\bibitem[{{Miller} \& {Owen}(2003)}]{MO03}
{Miller}, N.~A. \& {Owen}, F.~N. 2003, \aj, 125, 2427

\bibitem[{{Moore} {et~al.}(1996){Moore}, {Katz}, {Lake}, {Dressler}, \&
  {Oemler}}]{Moore+96}
{Moore}, B., {Katz}, N., {Lake}, G., {Dressler}, A., \& {Oemler}, Jr., A. 1996,
  \nat, 379, 613

\bibitem[{{Moran} {et~al.}(2005){Moran}, {Ellis}, {Treu}, {Smail}, {Dressler},
  {Coil}, \& {Smith}}]{Moran+05}
{Moran}, S.~M., {Ellis}, R.~S., {Treu}, T., {et~al.} 2005, \apj, 634, 977

\bibitem[{{Moran} {et~al.}(2007){Moran}, {Ellis}, {Treu}, {Smith}, {Rich}, \&
  {Smail}}]{Moran+07}
{Moran}, S.~M., {Ellis}, R.~S., {Treu}, T., {et~al.} 2007, \apj, 671, 1503

\bibitem[{{Navarro} {et~al.}(1997){Navarro}, {Frenk}, \& {White}}]{NFW97}
{Navarro}, J.~F., {Frenk}, C.~S., \& {White}, S. D.~M. 1997, \apj, 490, 493

\bibitem[{{Neistein} {et~al.}(2006){Neistein}, {van den Bosch}, \&
  {Dekel}}]{NvdBD06}
{Neistein}, E., {van den Bosch}, F.~C., \& {Dekel}, A. 2006, \mnras, 372, 933

\bibitem[{{Netzer} {et~al.}(2007){Netzer}, {Lutz}, {Schweitzer}, {Contursi},
  {Sturm}, {Tacconi}, {Veilleux}, {Kim}, {Rupke}, {Baker}, {Dasyra},
  {Mazzarella}, \& {Lord}}]{Netzer+07}
{Netzer}, H., {Lutz}, D., {Schweitzer}, M., {et~al.} 2007, \apj, 666, 806

\bibitem[{{Nordon} {et~al.}(2011){Nordon}, {Lutz}, {Berta}, {Wuyts},
  {Magnelli}, {Altieri}, {Andreani}, {Aussel}, {Bongiovanni}, {Cepa},
  {Cimatti}, {Daddi}, {Elbaz}, {Fadda}, {Forster Schreiber}, {Genzel},
  {Lagache}, {Maiolino}, {Perez Garcia}, {Poglitsch}, {Popesso}, {Pozzi},
  {Rosario}, {Saintonge}, {Sanchez-Portal}, {Santini}, {Sturm}, {Tacconi},
  {Valtchanov}, \& {Yan}}]{Nordon+11}
{Nordon}, R., {Lutz}, D., {Berta}, S., {et~al.} 2011, 1106.1186

\bibitem[{{Park} \& {Hwang}(2009)}]{Park+09}
{Park}, C. \& {Hwang}, H.~S. 2009, \apj, 699, 1595

\bibitem[{{Patel} {et~al.}(2009){Patel}, {Holden}, {Kelson}, {Illingworth}, \&
  {Franx}}]{Patel+09}
{Patel}, S.~G., {Holden}, B.~P., {Kelson}, D.~D., {Illingworth}, G.~D., \&
  {Franx}, M. 2009, \apjl, 705, L67

\bibitem[{{Peng} {et~al.}(2010){Peng}, {Lilly}, {Kova{\v c}}, {Bolzonella},
  {Pozzetti}, {Renzini}, {Zamorani}, {Ilbert}, {Knobel}, {Iovino}, {Maier},
  {Cucciati}, {Tasca}, {Carollo}, {Silverman}, {Kampczyk}, {de Ravel},
  {Sanders}, {Scoville}, {Contini}, {Mainieri}, {Scodeggio}, {Kneib}, {Le
  F{\`e}vre}, {Bardelli}, {Bongiorno}, {Caputi}, {Coppa}, {de la Torre},
  {Franzetti}, {Garilli}, {Lamareille}, {Le Borgne}, {Le Brun}, {Mignoli},
  {Perez Montero}, {Pello}, {Ricciardelli}, {Tanaka}, {Tresse}, {Vergani},
  {Welikala}, {Zucca}, {Oesch}, {Abbas}, {Barnes}, {Bordoloi}, {Bottini},
  {Cappi}, {Cassata}, {Cimatti}, {Fumana}, {Hasinger}, {Koekemoer},
  {Leauthaud}, {Maccagni}, {Marinoni}, {McCracken}, {Memeo}, {Meneux}, {Nair},
  {Porciani}, {Presotto}, \& {Scaramella}}]{Peng+10}
{Peng}, Y., {Lilly}, S.~J., {Kova{\v c}}, K., {et~al.} 2010, \apj, 721, 193

\bibitem[{{Pimbblet}(2003)}]{Pimbblet03}
{Pimbblet}, K.~A. 2003, \pasa, 20, 294

\bibitem[{{Pipino} \& {Matteucci}(2004)}]{PM04}
{Pipino}, A. \& {Matteucci}, F. 2004, \mnras, 347, 968

\bibitem[{{Poggianti} {et~al.}(2004){Poggianti}, {Bridges}, {Komiyama}, {Yagi},
  {Carter}, {Mobasher}, {Okamura}, \& {Kashikawa}}]{Poggianti+04}
{Poggianti}, B.~M., {Bridges}, T.~J., {Komiyama}, Y., {et~al.} 2004, \apj, 601,
  197

\bibitem[{{Poggianti} {et~al.}(1999){Poggianti}, {Smail}, {Dressler}, {Couch},
  {Barger}, {Butcher}, {Ellis}, \& {Oemler}}]{Poggianti+99}
{Poggianti}, B.~M., {Smail}, I., {Dressler}, A., {et~al.} 1999, \apj, 518, 576

\bibitem[{{Polletta} {et~al.}(2007){Polletta}, {Tajer}, {Maraschi},
  {Trinchieri}, {Lonsdale}, {Chiappetti}, {Andreon}, {Pierre}, {Le F{\`e}vre},
  {Zamorani}, {Maccagni}, {Garcet}, {Surdej}, {Franceschini}, {Alloin},
  {Shupe}, {Surace}, {Fang}, {Rowan-Robinson}, {Smith}, \&
  {Tresse}}]{Polletta+07}
{Polletta}, M., {Tajer}, M., {Maraschi}, L., {et~al.} 2007, \apj, 663, 81

\bibitem[{{Popesso} {et~al.}(2009){Popesso}, {Dickinson}, {Nonino}, {Vanzella},
  {Daddi}, {Fosbury}, {Kuntschner}, {Mainieri}, {Cristiani}, {Cesarsky},
  {Giavalisco}, {Renzini}, \& {GOODS Team}}]{Popesso+09}
{Popesso}, P., {Dickinson}, M., {Nonino}, M., {et~al.} 2009, \aap, 494, 443

\bibitem[{{Popesso} {et~al.}(2011){Popesso}, {Rodighiero}, {Saintonge},
  {Santini}, {Grazian}, {Lutz}, {Brusa}, \& {PEP Consortium}}]{Popesso+11}
{Popesso}, P., {Rodighiero}, G., {Saintonge}, A., {et~al.} 2011, 1104.1094

\bibitem[{{Porter} {et~al.}(2008){Porter}, {Raychaudhury}, {Pimbblet}, \&
  {Drinkwater}}]{Porter+08}
{Porter}, S.~C., {Raychaudhury}, S., {Pimbblet}, K.~A., \& {Drinkwater}, M.~J.
  2008, \mnras, 388, 1152

\bibitem[{{Postman} {et~al.}(2005){Postman}, {Franx}, {Cross}, {Holden},
  {Ford}, {Illingworth}, {Goto}, {Demarco}, {Rosati}, {Blakeslee}, {Tran},
  {Ben{\'{\i}}tez}, {Clampin}, {Hartig}, {Homeier}, {Ardila}, {Bartko},
  {Bouwens}, {Bradley}, {Broadhurst}, {Brown}, {Burrows}, {Cheng}, {Feldman},
  {Golimowski}, {Gronwall}, {Infante}, {Kimble}, {Krist}, {Lesser}, {Martel},
  {Mei}, {Menanteau}, {Meurer}, {Miley}, {Motta}, {Sirianni}, {Sparks}, {Tran},
  {Tsvetanov}, {White}, \& {Zheng}}]{Postman+05}
{Postman}, M., {Franx}, M., {Cross}, N.~J.~G., {et~al.} 2005, \apj, 623, 721

\bibitem[{{Prescott} {et~al.}(2006){Prescott}, {Impey}, {Cool}, \&
  {Scoville}}]{Prescott+06}
{Prescott}, M.~K.~M., {Impey}, C.~D., {Cool}, R.~J., \& {Scoville}, N.~Z. 2006,
  \apj, 644, 100

\bibitem[{{Rodighiero} {et~al.}(2010){Rodighiero}, {Cimatti}, {Gruppioni},
  {Popesso}, {Andreani}, {Altieri}, {Aussel}, {Berta}, {Bongiovanni},
  {Brisbin}, {Cava}, {Cepa}, {Daddi}, {Dominguez-Sanchez}, {Elbaz}, {Fontana},
  {F{\"o}rster Schreiber}, {Franceschini}, {Genzel}, {Grazian}, {Lutz},
  {Magdis}, {Magliocchetti}, {Magnelli}, {Maiolino}, {Mancini}, {Nordon},
  {Perez Garcia}, {Poglitsch}, {Santini}, {Sanchez-Portal}, {Pozzi},
  {Riguccini}, {Saintonge}, {Shao}, {Sturm}, {Tacconi}, {Valtchanov},
  {Wetzstein}, \& {Wieprecht}}]{Rodighiero+10}
{Rodighiero}, G., {Cimatti}, A., {Gruppioni}, C., {et~al.} 2010, \aap, 518, L25

\bibitem[{{Saintonge} {et~al.}(2008){Saintonge}, {Tran}, \& {Holden}}]{STH08}
{Saintonge}, A., {Tran}, K.-V.~H., \& {Holden}, B.~P. 2008, \apjl, 685, L113

\bibitem[{{Sanders} {et~al.}(2007){Sanders}, {Salvato}, {Aussel}, {Ilbert},
  {Scoville}, {Surace}, {Frayer}, {Sheth}, {Helou}, {Brooke}, {Bhattacharya},
  {Yan}, {Kartaltepe}, {Barnes}, {Blain}, {Calzetti}, {Capak}, {Carilli},
  {Carollo}, {Comastri}, {Daddi}, {Ellis}, {Elvis}, {Fall}, {Franceschini},
  {Giavalisco}, {Hasinger}, {Impey}, {Koekemoer}, {Le F{\`e}vre}, {Lilly},
  {Liu}, {McCracken}, {Mobasher}, {Renzini}, {Rich}, {Schinnerer}, {Shopbell},
  {Taniguchi}, {Thompson}, {Urry}, \& {Williams}}]{Sanders+07}
{Sanders}, D.~B., {Salvato}, M., {Aussel}, H., {et~al.} 2007, \apjs, 172, 86

\bibitem[{{Santini} {et~al.}(2009){Santini}, {Fontana}, {Grazian}, {Salimbeni},
  {Fiore}, {Fontanot}, {Boutsia}, {Castellano}, {Cristiani}, {de Santis},
  {Gallozzi}, {Giallongo}, {Menci}, {Nonino}, {Paris}, {Pentericci}, \&
  {Vanzella}}]{Santini+09}
{Santini}, P., {Fontana}, A., {Grazian}, A., {et~al.} 2009, \aap, 504, 751

\bibitem[{{Smith} {et~al.}(2005){Smith}, {Treu}, {Ellis}, {Moran}, \&
  {Dressler}}]{Smith+05}
{Smith}, G.~P., {Treu}, T., {Ellis}, R.~S., {Moran}, S.~M., \& {Dressler}, A.
  2005, \apj, 620, 78

\bibitem[{{Springel} \& {Farrar}(2007)}]{SF07}
{Springel}, V. \& {Farrar}, G.~R. 2007, \mnras, 380, 911

\bibitem[{{Temporin} {et~al.}(2009){Temporin}, {Duc}, {Ilbert}, \&
  {XMM-LSS/SWIRE collaboration}}]{TDI09}
{Temporin}, S., {Duc}, P., {Ilbert}, O., \& {XMM-LSS/SWIRE collaboration}.
  2009, Astronomische Nachrichten, 330, 915

\bibitem[{{Tran} {et~al.}(2007){Tran}, {Franx}, {Illingworth}, {van Dokkum},
  {Kelson}, {Blakeslee}, \& {Postman}}]{Tran+07}
{Tran}, K.-V.~H., {Franx}, M., {Illingworth}, G.~D., {et~al.} 2007, \apj, 661,
  750

\bibitem[{{Tran} {et~al.}(2009){Tran}, {Saintonge}, {Moustakas}, {Bai},
  {Gonzalez}, {Holden}, {Zaritsky}, \& {Kautsch}}]{Tran+09}
{Tran}, K.-V.~H., {Saintonge}, A., {Moustakas}, J., {et~al.} 2009, \apj, 705,
  809

\bibitem[{{Trump} {et~al.}(2007){Trump}, {Impey}, {McCarthy}, {Elvis},
  {Huchra}, {Brusa}, {Hasinger}, {Schinnerer}, {Capak}, {Lilly}, \&
  {Scoville}}]{Trump+07}
{Trump}, J.~R., {Impey}, C.~D., {McCarthy}, P.~J., {et~al.} 2007, \apjs, 172,
  383

\bibitem[{{Yahil} \& {Vidal}(1977)}]{YV77}
{Yahil}, A. \& {Vidal}, N.~V. 1977, \apj, 214, 347

\bibitem[{{Yee} {et~al.}(1996){Yee}, {Ellingson}, {Abraham}, {Gravel},
  {Carlberg}, {Smecker-Hane}, {Schade}, \& {Rigler}}]{Yee+96}
{Yee}, H.~K.~C., {Ellingson}, E., {Abraham}, R.~G., {et~al.} 1996, \apjs, 102,
  289

\bibitem[{{Zabludoff} \& {Mulchaey}(1998)}]{ZM98let}
{Zabludoff}, A.~I. \& {Mulchaey}, J.~S. 1998, \apjl, 498, L5

\end{thebibliography}

\end{document}